\documentstyle[12pt]{article}
\def \beq{\begin{equation}}
\def \dg{\Delta \Gamma}
\def \eeq{\end{equation}}
\def \gb{\Gamma(B^0)}
\def \gdc{\Gamma(D^+)}
\def \gdn{\Gamma(D^0)}
\def \glb{\Gamma(\Lambda_b)}
\def \glc{\Gamma(\Lambda_c)}
\def \lb{\Lambda_b}
\def \lc{\Lambda_c}
\def \psbd{|\Psi(0)|^2_{bd}}
\def \psbu{|\Psi(0)|^2_{bu}}
\def \psbub{|\Psi(0)|^2_{b \bar u}}
\def \pscd{|\Psi(0)|^2_{cd}}
\def \pscdb{|\Psi(0)|^2_{c \bar d}}

\def \sb{\Sigma_b}
\def \sbs{\Sigma_b^*}
\def \sc{\Sigma_c}
\def \scs{\Sigma_c^*}

\def \tb{\tau(B^0)}

\def \tlb{\tau(\Lambda_b)}

\begin{document}
\renewcommand{\thetable}{\Roman{table}}
\rightline{CERN-TH-96/24}
\rightline{EFI-96-03}
\rightline{hep-ph/9602265}
\rightline{February 1996}
\bigskip
\bigskip
\centerline{{\bf ENHANCEMENT OF THE $\Lambda_b$ DECAY RATE}
\footnote{To be submitted to Phys.~Lett.~B.}}
\bigskip
\centerline{\it Jonathan L. Rosner}
\medskip
\centerline{\it Div.~TH, CERN}
\centerline{\it 1211 CH Geneva 23, Switzerland}
\smallskip
\centerline{and}
\smallskip
\centerline{\it Enrico Fermi Institute and Department of Physics}
\centerline{\it University of Chicago, Chicago, IL 60637
\footnote{Permanent address.}}
\bigskip
\centerline{\bf ABSTRACT}
\medskip
\begin{quote}
The enhancement $\Delta \Gamma (\Lambda_b)$ of the $\Lambda_b$ decay rate due
to four-fermion processes of weak scattering and Pauli interference is
calculated within the quark model. An estimate of the relative $bu$ wave
function at zero separation, $|\Psi(0)|^2_{bu}$, is obtained in terms of the
$\Sigma_b^* - \Sigma_b$ hyperfine splitting, the $B^* - B$ hyperfine splitting,
and the $B$ meson decay constant $f_B$.  For $M(\Sigma_b^*) - M(\Sigma_b) = 56
\pm 16$ MeV, $M(B^*) - M(B) = 46$ MeV, and $f_B = 190 \pm 40$ MeV, we find
$\Delta \Gamma(\Lambda_b) = (0.025 \pm 0.013)$ ps$^{-1}$, to be compared with
the observed enhancement $\Gamma(\Lambda_b) - \Gamma(B^0) = 0.20 \pm 0.05$
ps$^{-1}$.  Even such a meager enhancement entails a value of
$|\Psi(0)|^2_{bu}$ considerably larger than the corresponding value of
$|\Psi(0)|^2_{cd}$ in the $\Lambda_c$ baryon. 
\end{quote}
\medskip
\leftline{PACS codes: 14.20.Mr, 14.20.Lq, 13.30.Eg, 12.39.Jh}
\newpage

The differences among lifetimes of particles containing heavy quarks are
expected to become smaller as the heavy quark mass increases and free-quark
estimates become more reliable.  Thus, although charmed particles have
lifetimes ranging from less than 0.1 ps for the $\Omega_c$ \cite{Oclife} to
greater than 1 ps for the $D^+$ \cite{PDG}, mesons and baryons containing $b$
quarks are expected to have lifetimes differing no more than a few percent
\cite{Bigia,Bigib,NS}.  For example, it is expected that the process, $b u \to
c d$ in the $\lb$ (``weak scattering''), when considered in conjunction with
the partially offsetting process $b d \to c \bar u d d$ (``Pauli
interference'') should lead to a small enhancement in the $\lb$ decay rate, so
that $\tlb = (0.9~{\rm to}~0.95) \tb$.  Some caution has been urged with regard
to the four-quark matrix element in these estimates \cite{MSpc}. In the present
Letter we present a new evaluation of this matrix element which confirms the
expected smallness of the enhancement.  We perform this evaluation using a
hyperfine splitting sensitive to the heavy quark -- light quark interaction,
which has recently become possible in $b$-flavored baryons as a result of a
measurement of the $\sbs - \sb$ splitting by the DELPHI Collaboration
\cite{Delphi}. 

The observed $\lb$ lifetime is $\tlb = 1.20 \pm 0.07$ ps, while the $B^0$ 
decays more slowly:  $\tb = (1.58 \pm 0.05)$ ps.  Here we have averaged a
compilation of world data \cite{Sharma} (for which $\tb = 1.18 \pm 0.07$ ps)
with a new value \cite{CDFtb} $\tb = 1.33 \pm 0.16 \pm 0.07$ ps.  The ratio of
these two quantities is $\tlb/\tb = 0.76 \pm 0.05$, indicating an enhancement
of the $\lb$ decay rate beyond the magnitude of usual estimates. 

In the present Letter we find that, in spite of a large wave function for the
$b u$ pair in the initial baryon, which we denote by $\psbu$, only $(13 \pm
7)\%$ of the needed enhancement of the $\lb$ decay rate can be explained in
terms of the effects of the four-fermion matrix element. (Isospin symmetry then
dictates $\psbd = \psbu$.)  If we assume wave functions are similar in all
baryons with a single $b$ quark and two nonstrange quarks, this quantity can be
related to the hyperfine splitting $M(\sbs) - M(\sb)$, for which the DELPHI
Collaboration at LEP \cite{Delphi} has recently quoted a large value of $56 \pm
16$ MeV.  We estimate the effect of gluon exchange by performing a similar
calculation for $B$ mesons, relating the $B^* - B$ splitting to the $B$ meson
decay constant and taking account of differing spin and hyperfine factors in
the meson and baryon systems. 

A relation for the enhancement of the $\lc$ decay rate due to the weak
scattering process $c d \to s u$ was first pointed out in Ref.~\cite{BLS}.  At
the same order in heavy quark mass, one must also take account of Pauli
interference (interference between identical quarks in the final state)
\cite{VS,Gub}. Thus, for the $\lb$, one considers not only the process $b u \to
c d$ (involving matrix elements between $\lb$ states of $(\bar b b)(\bar u u)$
operators), but also those processes involving matrix elements of $(\bar b
b)(\bar d d)$ operators) which contribute to interference.  The net result of
four-quark operators in the $\lb$ is an enhancement of the decay rate by an
amount (see, e.g., Refs.~\cite{VS,Gub,NS}) 

\beq \label{eqn:dg}
\dg(\lb) = \frac{G_F^2}{2 \pi} \psbu |V_{ud}|^2
|V_{cb}|^2 m_b^2(1-x)^2 [c_-^2 - (1+x)c_+(c_- - c_+/2)]~~~.
\eeq
Here we have neglected light-quark masses; $x \equiv m_c^2/m_b^2$, while $c_-$
and $c_+ = (c_-)^{-1/2}$ are the short-distance QCD enhancement and suppression
factors for quarks in a color antitriplet and sextet, respectively:
\cite{QCDenh}: 
\beq
c_- = \left[ \frac{\alpha_s(m_c^2)}{\alpha_s(M_W^2)} \right ]^\gamma~~~,~~~
\gamma = \frac{12}{33 - 2 n_F}~~~,
\eeq
with $n_F = 5$ the number of active quark flavors between $m_b$ and $M_W$.
The $c_-^2$ term in square brackets reflects the weak scattering process $bu
\to cd \to bu$, while the remaining terms arise from destructive interference
between the two intermediate $d$ quarks in the process $bd \to c \bar u dd \to
bd$. 

Taking the strong interaction scale in the modified-minimal-subtraction scheme
for four quark flavors to be \cite{KMRR} $\Lambda_{\overline{\rm MS}}^{(4)} =
200$ MeV, we find $\alpha_s(m_b^2) = 0.193$ and $\alpha_s(M_W^2) = 0.114$, and
hence $c_- = 1.32,~c_+ = 0.87$.  An estimate of $\psbu$ is then needed.  We
find it by comparing hyperfine splittings in mesons and baryons, under the
assumption that the strength of the one-gluon exchange term is the same for the
light quark -- heavy quark pair in each system. 

Our result, which we shall explain in more detail presently, is
\beq \label{eqn:ps}
\psbu = 2 \cdot \frac{2}{3} \cdot \frac{M(\sbs) - M(\sb)}{M(B^*) - M(B)}
\cdot \frac{M_B f_B^2}{12}~~~,
\eeq
where the first factor relates to color, the second to spin, and the last term
is the nonrelativistic estimate of the $b \bar u$ wave function in the $B$
meson \cite{wf}. (Here one may use the spin-averaged value of vector and
pseudoscalar masses for $M_M$.)  With the DELPHI value of $M(\sbs) - M(\sb)$,
the $B^* - B$ splitting of 46 MeV \cite{PDG}, and the estimate \cite{fm,lat}
$f_B = 190 \pm 40$ MeV, we obtain $\psbu = (2.6 \pm 1.3) \times 10^{-2}$
GeV$^3$. This is to be compared with $\psbub = M_B f_B^2/12 = (1.6 \pm 0.7)
\times 10^{-2}$ GeV$^3$ for the $B$ meson.  Our assumption of equal values of
$\alpha_s$ governing the hyperfine interaction of the light -- heavy pair in
the meson and baryon must be viewed cautiously in this light.  It is possible
that $\alpha_s$ in the more compact baryonic system (since the wave function
appears to be larger) is smaller than in the meson, which would further enhance
the estimate of $\psbu$.

Several previous estimates of the four-quark matrix element (see, e.g.,
Refs.~\cite{Bigib,BLS,VS,BS}) utilized the hyperfine splitting between the
$\Lambda_b$ and the $\Sigma_b$ (or the corresponding charmed states). Since
this splitting involves the interaction of two light quarks as well as one
light and one heavy quark, it necessarily involves a statement about light
quark masses, as well as about the relation between the relative wave function
of two light quarks and that of a light and a heavy quark.  Our approach avoids
such assumptions. 

In the relation (\ref{eqn:dg}) we now neglect $\sin \theta_c$ (setting $V_{ud}
=1$), and choose $m_b = 5.1$ GeV, $m_c = 1.7$ GeV, and $|V_{cb}| = 0.040 \pm
0.003$. We then find 
\beq
\dg(\lb) = 0.025 \pm 0.013~{\rm ps}^{-1}~~~.
\eeq

The decay rates of the $B^0$ and $\lb$ are $\gb = 0.63 \pm 0.02$ ps$^{-1}$ and
$\glb = 0.83 \pm 0.05$ ps$^{-1}$, differing by $\dg(\lb) = 0.20 \pm 0.05$
ps$^{-1}$. The four-quark processes noted above can explain only $(13 \pm 7)\%$
of this difference, leading to an enhancement of only $(4 \pm 2)\%$ of the
total $\lb$ decay rate in contrast with the needed enhancement of $(32 \pm
8)\%$.

We now give some details of the calculation and see how well it does for the
$\lc$.

The hyperfine interaction in a meson $M_{i \bar j}$ composed of a quark $i$ and
an antiquark $\bar j$ leads to a mass shift \cite{DGG}
\beq \label{eqn:mhfs}
\Delta M(M_{i \bar j}) = \frac{32 \pi}{9} \alpha_s \frac{\langle \hat{s}_i
\cdot \hat{s}_j \rangle}{m_i m_j} |\Psi(0)|^2~~~,
\eeq
where $\hat s_i$ refers to a quark spin operator.  The corresponding result for
a baryon $B_{ijk}$ is 
\beq \label{eqn:bhfs}
\Delta M(B_{ijk}) = \frac{16 \pi}{9} \alpha_s \sum_{i > j} \frac{\langle
\hat{s}_i \cdot \hat{s}_j \rangle}{m_i m_j} |\Psi(0)|_{ij}^2~~~.
\eeq
The relative factor of 2 arises from the different color factors for a
quark and antiquark in a meson (a triplet and antitriplet making a singlet)
and two quarks in a baryon (two triplets making an antitriplet).  This factor
of 2 is the first term in Eq.~(\ref{eqn:ps}).

We take the ratio of hyperfine splittings in mesons and baryons so that light
quark masses and values of $\alpha_s$ cancel out.  We thus assume that (a)
effective light quark masses in mesons and baryons are equal (borne out at the
20\% level by phenomenological fits to meson and baryon spectra \cite{DGG}),
and (b) the values of $\alpha_s$ governing the corresponding hyperfine
splittings are similar (reasonable since in both mesons and baryons one is
concerned with systems of one light and one heavy quark).

We must isolate a baryon mass shift sensitive to the interaction between a
light quark and a heavy one.  The $\sbs - \sb$ splitting is the appropriate
quantity.  In both $\sb$ and $\sbs$, the light quarks are coupled up to
spin 1.  The splitting then depends purely on the light quark -- heavy quark
interaction.

The wave function between a light quark and a heavy one is assumed, as
mentioned, to be identical in the $\lb$ and in the $\sb - \sbs$ system.  The
two light quarks are coupled up to zero spin in the $\lb$, and hence have
zero net hyperfine intreraction with the heavy quark, while the hyperfine
interaction between the light quarks and the heavy one averages to zero
if we take the spin-weighted average of the $\sb$ and the $\sbs$.

The value of $\langle \hat{s}_Q \cdot \hat{s}_{\bar q} \rangle$ is $(1/4,-3/4)$
for a $(^3S_1,~^1S_0)$ $Q \bar q$ meson, where $Q$ and $q$ are the heavy and
light quark.  For a baryon $Qqq$ with $S_{qq} = 1$, one has $\langle \hat{s}_Q
\cdot \hat{s}_q \rangle = (1/4,~-1/2)$ for states with total spin (3/2,~1/2).
Thus the difference in $\hat{s}_i \cdot \hat{s}_j$ for the $\sbs - \sb$
splitting (counting a factor of 2 for the two light quarks in the baryons) is
3/2 that for the $B^* - B$ splitting. The factor of 2/3 in Eq.~(\ref{eqn:ps})
compensates for this ratio. 

The relation \cite{wf} $|\Psi(0)|^2 = M_M f_M^2 /12$ for the square of the
wave function of a $Q \bar q$ meson probably has important corrections of order
$1/m_Q$, if lattice calculations are any guide \cite{lat}.  These are ignored
in the present discussion.  They are likely to be more important when we
apply the present method to charmed states.

The corresponding calculation for charmed particles makes use of the
following inputs.

{\it 1. The $D$ meson decay constant} was taken \cite{fm} to be $f_D =
240 \pm 40$ MeV, leading (with $M_D = 1973$ MeV) to $\pscdb = (0.95 \pm 0.32)
\times 10^{-2}$ GeV$^3$.

{\it 2. The $D^* - D$ splitting} is assumed to be 141 MeV (the average for
charged and neutral states \cite{PDG}).

{\it 3. Charmed baryon masses} are taken to be $M(\sc) = 2453$ MeV \cite{PDG}
and $M(\scs) = 2530 \pm 7$ MeV \cite{Ammosov}.

{\it 4. The strong fine-structure-constant} at $m_c^2$ is taken to be
$\alpha_s(m_c^2) = 0.289$, consistent with the QCD scale mentioned above,
leading to $c_- = 1.60,~c_+ = 0.79$.

{\it 5. The strange quark mass} is taken to have a typical constituent-quark
value, $m_s = 0.5$ GeV.  We continue to neglect $u$ and $d$-quark masses for
simplicity.

{\it 6. The CKM factors} in Eq.~(\ref{eqn:dg}) undergo the replacements
$|V_{ud}|^2 |V_{cb}|^2 \to |V_{cs}|^2 |V_{ud}|^2$, which we approximate by
1 (again neglecting $\sin \theta_c$).

The resulting matrix element $\pscd = (0.69 \pm 0.24) \times 10^{-2}$ GeV$^3$
is consistent with an estimate in Ref.~\cite{BLS} of $\pscd = 0.74 \times
10^{-2}$ GeV$^3$. In that work, the value of $\pscd$ was assumed to be the same
as for two light quarks in the charmed baryon, and was estimated using the
observed $\sc - \lc$ splitting.  It was also necessary to assume a specific
value of $\alpha_s = 0.58$ in the hyperfine interaction expression (6), which
we do not do here. Our value of $c_-$ is smaller than assumed in
Ref.~\cite{BLS} and we take account of destructive interference, leading to a
smaller result for $\dg(\lc)$. 

The results for systems with $c$ and $b$ quarks based on our method are
summarized in Table I.  Several remarks can be made.

\renewcommand{\arraystretch}{1.3}
\begin{table}
\caption{Comparison of predicted squares of wave functions and decay rate
enhancements for $\lc$ and $\lb$.}
\begin{center}
\begin{tabular}{c c c} \hline
Quantity (units) & Charm & Beauty \\ \hline
$f_M$ (MeV)   &  $240 \pm 40$   &   $190 \pm 40$ \\
$|\Psi(0)|^2_{Q \bar q}~(10^{-2}$ GeV$^3$) & $0.95 \pm 0.32$
  & $1.6 \pm 0.7$ \\
$M(^3S_1) - M(^1S_0)$ (MeV) & 141 & 46 \\
$M(\Sigma^*) - M(\Sigma)$ (MeV) & $77 \pm 7$ & $56 \pm 16$ \\
$|\Psi(0)|^2_{Q q}~(10^{-2}$ GeV$^3$) & $0.69 \pm 0.24$
  & $2.6 \pm 1.3$ \\
$c_-$ & 1.60 & 1.32 \\
$c_+$ & 0.79 & 0.87 \\
$c_-^2 - (1+x)c_+(c_- - c_+^2/2)$ & 1.52 & 0.88 \\
$\dg(\Lambda_Q)$ (ps$^{-1}$) & $0.8\pm 0.3$ & $0.025 \pm 0.013$ \\ \hline
\end{tabular}
\end{center}
\end{table}

(a) The difference between the central values of $|\Psi(0)|^2_{Q \bar q}$ for
charm and beauty reflects the likely importance of $1/m_Q$ corrections (see,
e.g., Ref.~\cite{lat}), or -- in the language of the quark model -- of reduced
mass effects. 

(b) The $\scs - \sc$ hyperfine splitting used in this calculation is based on
one claim for observation of the $\scs$ \cite{Ammosov}, which requires
confirmation. 

(c) The value of $\psbu$ is somewhat large in comparison with the others for
light-heavy systems.  It would be helpful to verify the large hyperfine
splitting between $\sbs$ and $\sb$ claimed by the DELPHI Collaboration
\cite{Delphi}.  The ratio of hyperfine splittings for charmed and beauty mesons
is approximately 3:1, as expected if these splittings scale as $1/m_Q$.  In
contrast, the corresponding ratio for baryons is considerably smaller,
indicating a violation of $1/m_Q$ scaling. 

(d) The enhancement of the $\lc$ decay rate is quite modest. With $\glc \approx
5$ ps$^{-1}$, to be compared with $\gdn \approx 2.4$ ps$^{-1}$ and $\gdc
\approx 1$ ps$^{-1}$, one seeks an enhancement of at least $\glc - \gdn \approx
2.6$ ps$^{-1}$. If the enhancements $\dg(\Lambda_Q)$ in Table I were about a
factor of 4 larger, we could accommodate both the $\lc$ and $\lb$ decay rates,
but this is not consistent with our estimates of the matrix elements and their
effects on decay rates.  In particular, the effect of Pauli interference is to
cut the na\"{\i}ve estimate of the enhancement due to weak scattering alone
\cite{BLS} by roughly a factor of 2.  Hybrid logarithms \cite{VS}, not
considered here, have a relatively modest effect, leading if anything to
further suppression of the enhancement for $\lb$ decay \cite{NS}. 

To summarize, we have used the hyperfine splitting between $\sbs$ and $\sb$
claimed by the DELPHI Collaboration \cite{Delphi} to estimate the overlap of
quark wave functions between the $b$ quark and the light quarks in the
$\Lambda_b$, and hence to estimate the effect of four-quark operators on its
decay rate.  Even though the matrix element $\psbu = \psbd$ deduced from the
DELPHI result is quite large on the scale of those for heavy-light systems, one
can only account for $(13 \pm 7)\%$ of the difference between the $\lb$ and
$B^0$ decay rates, or an enhancement of $(4 \pm 2)\%$ of the $\lb$ decay rate.
A similar approach also falls short of accounting for the corresponding
enhancement for the $\lc$ decay rate.  If the enhanced $\lb$ decay rate is
borne out by further data, we can only speculate that strong final-state
interactions which cannot be anticipated on the basis of perturbative QCD must
play a role even at the rather high mass of the $\lb$. 

I wish to thank G. Martinelli, M. Neubert, C. T. Sachrajda, M. Shifman, and N.
G. Uraltsev for fruitful discussions, and the Physics Department of the
Technion for its hospitality during the initial stages of this investigation. 
This work was supported in part by the United States -- Israel Binational
Science Foundation under Research Grant Agreement 94-00253/1 and by the United
States Department of Energy under Contract No. DE FG02 90ER40560. 

% Journal and other miscellaneous abbreviations for references
% Phys. Lett. B style
\def \ajp#1#2#3{Am.~J.~Phys.~{\bf#1} (#3) #2}
\def \apny#1#2#3{Ann.~Phys.~(N.Y.) {\bf#1} (#3) #2}
\def \app#1#2#3{Acta Phys.~Polonica {\bf#1} (#3) #2 }
\def \arnps#1#2#3{Ann.~Rev.~Nucl.~Part.~Sci.~{\bf#1} (#3) #2}
\def \cmp#1#2#3{Commun.~Math.~Phys.~{\bf#1} (#3) #2}
\def \cmts#1#2#3{Comments on Nucl.~Part.~Phys.~{\bf#1} (#3) #2}
\def \cn{Collaboration}
\def \corn93{{\it Lepton and Photon Interactions:  XVI International Symposium,
Ithaca, NY August 1993}, AIP Conference Proceedings No.~302, ed.~by P. Drell
and D. Rubin (AIP, New York, 1994)}
\def \cp89{{\it CP Violation,} edited by C. Jarlskog (World Scientific,
Singapore, 1989)}
\def \dpff{{\it The Fermilab Meeting -- DPF 92} (7th Meeting of the American
Physical Society Division of Particles and Fields), 10--14 November 1992,
ed. by C. H. Albright \ite~(World Scientific, Singapore, 1993)}
\def \dpf94{DPF 94 Meeting, Albuquerque, NM, Aug.~2--6, 1994}
\def \efi{Enrico Fermi Institute Report No. EFI}
\def \el#1#2#3{Europhys.~Lett.~{\bf#1} (#3) #2}
\def \f79{{\it Proceedings of the 1979 International Symposium on Lepton and
Photon Interactions at High Energies,} Fermilab, August 23-29, 1979, ed.~by
T. B. W. Kirk and H. D. I. Abarbanel (Fermi National Accelerator Laboratory,
Batavia, IL, 1979}
\def \hb87{{\it Proceeding of the 1987 International Symposium on Lepton and
Photon Interactions at High Energies,} Hamburg, 1987, ed.~by W. Bartel
and R. R\"uckl (Nucl. Phys. B, Proc. Suppl., vol. 3) (North-Holland,
Amsterdam, 1988)}
\def \ib{{\it ibid.}~}
\def \ibj#1#2#3{~{\bf#1} (#3) #2}
\def \ichep72{{\it Proceedings of the XVI International Conference on High
Energy Physics}, Chicago and Batavia, Illinois, Sept. 6--13, 1972,
edited by J. D. Jackson, A. Roberts, and R. Donaldson (Fermilab, Batavia,
IL, 1972)}
\def \ijmpa#1#2#3{Int.~J.~Mod.~Phys.~A {\bf#1} (#3) #2}
\def \ite{{\it et al.}}
\def \jmp#1#2#3{J.~Math.~Phys.~{\bf#1} (#3) #2}
\def \jpg#1#2#3{J.~Phys.~G {\bf#1} (#3) #2}
\def \lkl87{{\it Selected Topics in Electroweak Interactions} (Proceedings of
the Second Lake Louise Institute on New Frontiers in Particle Physics, 15--21
February, 1987), edited by J. M. Cameron \ite~(World Scientific, Singapore,
1987)}
\def \ky85{{\it Proceedings of the International Symposium on Lepton and
Photon Interactions at High Energy,} Kyoto, Aug.~19-24, 1985, edited by M.
Konuma and K. Takahashi (Kyoto Univ., Kyoto, 1985)}
\def \mpla#1#2#3{Mod.~Phys.~Lett.~A {\bf#1} (#3) #2}
\def \nc#1#2#3{Nuovo Cim.~{\bf#1} (#3) #2}
\def \np#1#2#3{Nucl.~Phys.~{\bf#1} (#3) #2}
\def \pisma#1#2#3#4{Pis'ma Zh.~Eksp.~Teor.~Fiz.~{\bf#1} (#3) #2 [JETP Lett.
{\bf#1} (#3) #4]}
\def \pl#1#2#3{Phys.~Lett.~{\bf#1} (#3) #2}
\def \plb#1#2#3{Phys.~Lett.~B {\bf#1} (#3) #2}
\def \pr#1#2#3{Phys.~Rev.~{\bf#1} (#3) #2}
\def \pra#1#2#3{Phys.~Rev.~A {\bf#1} (#3) #2}
\def \prd#1#2#3{Phys.~Rev.~D {\bf#1} (#3) #2}
\def \prl#1#2#3{Phys.~Rev.~Lett.~{\bf#1} (#3) #2}
\def \prp#1#2#3{Phys.~Rep.~{\bf#1} (#3) #2}
\def \ptp#1#2#3{Prog.~Theor.~Phys.~{\bf#1} (#3) #2}
\def \rmp#1#2#3{Rev.~Mod.~Phys.~{\bf#1} (#3) #2}
\def \rp#1{~~~~~\ldots\ldots{\rm rp~}{#1}~~~~~}
\def \si90{25th International Conference on High Energy Physics, Singapore,
Aug. 2-8, 1990}
\def \slc87{{\it Proceedings of the Salt Lake City Meeting} (Division of
Particles and Fields, American Physical Society, Salt Lake City, Utah, 1987),
ed.~by C. DeTar and J. S. Ball (World Scientific, Singapore, 1987)}
\def \slac89{{\it Proceedings of the XIVth International Symposium on
Lepton and Photon Interactions,} Stanford, California, 1989, edited by M.
Riordan (World Scientific, Singapore, 1990)}
\def \smass82{{\it Proceedings of the 1982 DPF Summer Study on Elementary
Particle Physics and Future Facilities}, Snowmass, Colorado, edited by R.
Donaldson, R. Gustafson, and F. Paige (World Scientific, Singapore, 1982)}
\def \smass90{{\it Research Directions for the Decade} (Proceedings of the
1990 Summer Study on High Energy Physics, June 25 -- July 13, Snowmass,
Colorado), edited by E. L. Berger (World Scientific, Singapore, 1992)}
\def \stone{{\it B Decays}, edited by S. Stone (World Scientific, Singapore,
1994)}
\def \tasi90{{\it Testing the Standard Model} (Proceedings of the 1990
Theoretical Advanced Study Institute in Elementary Particle Physics, Boulder,
Colorado, 3--27 June, 1990), edited by M. Cveti\v{c} and P. Langacker
(World Scientific, Singapore, 1991)}
\def \yaf#1#2#3#4{Yad.~Fiz.~{\bf#1} (#3) #2 [Sov.~J.~Nucl.~Phys.~{\bf #1} (#3)
#4]}
\def \zhetf#1#2#3#4#5#6{Zh.~Eksp.~Teor.~Fiz.~{\bf #1} (#3) #2 [Sov.~Phys. -
JETP {\bf #4} (#6) #5]}
\def \zpc#1#2#3{Zeit.~Phys.~C {\bf#1} (#3) #2}

\end{document}